\documentclass[reprint,prl,twocolumn,superscriptaddress,floatfix,preprintnumbers]{revtex4-1}

\usepackage[utf8]{inputenc}
\usepackage[english]{babel}
\usepackage{amsmath,amssymb,mathtools}
\usepackage{graphicx}
\usepackage{braket}
\usepackage{booktabs}
\usepackage{xcolor}
\usepackage[hidelinks]{hyperref}
\usepackage{cleveref}
\usepackage{float}
\usepackage{tikz}
\usetikzlibrary{arrows.meta,decorations.markings}

\newcommand{\ba}{\begin{eqnarray}}
\newcommand{\ea}{\end{eqnarray}}
\newcommand{\nn}{\nonumber}
\newcommand{\be}{\begin{equation}}
\newcommand{\ee}{\end{equation}}
\newcommand{\vecp}[1]{\vec{#1}\hspace{1.2pt}}
\newcommand{\vecpp}[1]{\vec{#1}\hspace{2.5pt}}

\newcommand{\GeV}{\mathrm{GeV}}
\newcommand{\fm}{\mathrm{fm}}
\newcommand{\MeV}{\mathrm{MeV}}

\newcommand{\Lweak}{\mathcal{L}^{\Delta S=1}}

\usepackage{slashed}
\newcommand{\innovation}{Collaborative Innovation Center of Quantum Matter, Beijing 100871, China}
\newcommand{\chep}{Center for High Energy Physics, Peking University, Beijing 100871, China}
\newcommand{\pkuphy}{School of Physics, Peking University, Beijing 100871, China}
\newcommand{\Uconn}{Department of Physics, University of Connecticut, Storrs, CT 06269, USA}
\newcommand{\SCNT}{Southern Center for Nuclear-Science Theory (SCNT), Institute of Modern Physics, Chinese Academy of Sciences, Guangdong 516000, China}
\newcommand{\Soton}{School of Physics and Astronomy, University of Southampton, Southampton, SO17 1BJ, UK}

\makeatletter
\def\pre@bibdata{}
\def\@bibdataout@init{}
\def\@bibdataout@rev{}
\makeatother

\begin{document}

\title{Emerging understanding of the rare kaon decay $K^+\to\pi^+\ell^+\ell^-$ from lattice QCD}

\author{Zhao-Long~Zhang}\affiliation{\pkuphy}
\author{Xu~Feng}\affiliation{\pkuphy}\affiliation{\innovation}\affiliation{\chep}\affiliation{\SCNT}
\author{Lu-Chang~Jin}\affiliation{\Uconn}
\author{Bo-Lin~Li}\affiliation{\pkuphy}
\author{Chuan~Liu}\affiliation{\pkuphy}\affiliation{\innovation}\affiliation{\chep}
\author{Christopher~T.~Sachrajda}\affiliation{\Soton}
\date{\today}

\begin{abstract}

Rare kaon decays provide sensitive probes of the Standard Model and potential new physics. For $K^+\to\pi^+\ell^+\ell^-$, the experimental decay rate determines only the squared vector form factor $|V(z)|^2$, leaving its sign undetermined. Previous lattice QCD calculations have either used unphysical pion masses or suffered from uncertainties too large to resolve this sign. Here we present a first-principles calculation using two domain-wall fermion ensembles with near-physical pion masses. We develop a computational strategy that establishes the ultraviolet finiteness and continuum equivalence of local and conserved vector currents, substantially simplifying the calculation of nonlocal matrix elements. Four independent subtraction schemes provide nontrivial consistency checks, while the one-end trick reduces noise from quark-self-loop contributions. We determine $V(z)$ at six spacelike and timelike momentum transfers. Fits consistently yield $V(0)<0$, with a statistical significance of $5.8\sigma$--$7.9\sigma$ at $M_{\eta_c}=1.1,\mathrm{GeV}$. After extrapolation to the physical charm-quark mass, the negative sign persists across all fit forms, with a significance of $2.0\sigma$--$3.4\sigma$. Our results provide direct first-principles evidence favoring the negative-sign solution and establish a framework for more precise lattice QCD determinations.

\end{abstract}

\maketitle

\paragraph{Introduction}

Rare kaon decays provide sensitive probes of the Standard Model and potential new physics. In particular, flavor-changing neutral-current (FCNC) kaon decays are strongly suppressed within the Standard Model, making them sensitive to small contributions from physics beyond the Standard Model~\cite{Gaillard:1974hs,Littenberg:1989ix,Littenberg:1993qv}. They therefore complement direct searches for new particles at high energies with precision tests at low energies~\cite{Littenberg:1993qv,DAmbrosio:2003hzz}.

Unlike the ultra-rare decay $K\to\pi\nu\bar{\nu}$, which is dominated by short-distance physics, the $K\to\pi\ell^+\ell^-$ decay amplitude
receives significant long-distance contributions through the nonlocal product of the electromagnetic current and the weak Hamiltonian, with the two vertices separated by distances of order $1/\Lambda_{\rm QCD}$. 
The resulting hadronic contribution is encoded in a one-photon-exchange form factor $V(z)$, where $z=q^2/M_K^2$ is the normalized dilepton invariant mass squared~\cite{DAmbrosio:2019xph,Ecker:1987qi}.

We focus on the charged mode $K^+\to\pi^+\ell^+\ell^-$, for which the experimental measurements are more precise and the amplitude structure is simpler than for the neutral modes. 
In particular, the charged mode does not involve the additional $\pi^0$-loop contribution with the insertion of the electromagnetic current that is present in the neutral modes.
Over the past several decades, substantial experimental efforts have been devoted to precision measurements of these rare decays. The $K^+\to\pi^+e^+e^-$ channel has been studied by CERN PS~\cite{Bloch:1974ua}, E777~\cite{Alliegro:1992pp}, E865~\cite{Appel:1999}, and NA48/2~\cite{Batley:2009aa}, while the $K^+\to\pi^+\mu^+\mu^-$ channel has been investigated by E787~\cite{E787:1997ohp}, E865~\cite{E865Muon:2000}, HyperCP (E871)~\cite{HyperCP:2001ebl}, NA48/2~\cite{Batley:2011zz}, and, more recently, NA62~\cite{NA62:2022qes}.

Since the decay rate is proportional to $|V(z)|^2$, it does not determine the sign of $V(z)$ directly.
To extract the form factor from the experimental spectra, one needs additional phenomenological input from the $K^+\to3\pi\to\pi^+\gamma^\ast$ contribution, denoted by $V^{\pi\pi}(z)$ and parametrized by 
$V(z)=a_++b_+z+V^{\pi\pi}(z)$~\cite{DAmbrosio:1998gur,DAmbrosio:2018ytt,Bernard:2025xyn,Ecker:1987hd,Friot:2004yr,Kubis:2010mp,Bruno:1992za}. Consequently, the inferred sign of $V(z)$, as well as the fitted values of $a_+$ and $b_+$, 
depends sensitively on the assumed $V^{\pi\pi}(z)$. Fits to the experimental spectra 
can admit two local minima with substantially different values of $(a_+,b_+)$, corresponding to opposite signs of $V(z)$ near $z=0$.

This ambiguity is evident in the analyses of the experimental data. For the $K^+\to\pi^+e^+e^-$ mode, 
an analysis of the E865 data finds a better fit for the branch with negative $a_+$ and $b_+$, with $\chi^2/{\rm dof}=11.6/18$,
while the positive branch gives $\chi^2/{\rm dof}=42.0/18$~\cite{DAmbrosio:2018ytt}. 
An analysis of the NA48/2 data, however, finds a less pronounced separation between the two branches: the negative and positive solutions give $\chi^2/{\rm dof}=33.0/19$ and $28.3/19$, respectively, while yielding substantially different parameter pairs, $(a_+,b_+)=(-0.585(13),-0.779(56))$ and $(0.491(13),1.691(59))$~\cite{DAmbrosio:2018ytt}. More recently, the precise $K^+\to\pi^+\mu^+\mu^-$ measurement by NA62 finds similarly good fits for both branches, with $\chi^2/{\rm dof}=45.1/48$ and $56.4/48$, respectively, despite markedly different parameter values, $(a_+,b_+)=(-0.575(12),-0.722(40))$ and $(0.373(12),2.017(40))$~\cite{NA62:2022qes}.
Although the negative branch is often adopted in phenomenological analyses, this choice is guided in part by expectations from chiral perturbation theory that $b_+$ should be subleading compared with $a_+$~\cite{DAmbrosio:2018ytt}, rather than being determined solely by the experimental spectra. Moreover, even for this branch, the fitted values typically satisfy $|b_+/a_+|\sim 1$, indicating that the expected hierarchy between $b_+$ and $a_+$ is not clearly borne out by the data.

Lattice QCD provides a first-principles approach to determine the form factor $V(z)$ directly from the underlying strong-interaction dynamics. Relevant studies date back more than a decade, with methods developed and exploratory numerical calculations performed for both $K\to\pi\ell^+\ell^-$ and $K\to\pi\nu\bar{\nu}$~\cite{Christ:2015aha,Christ:2016eae,Christ:2016mmq,Bai:2017fkh,Bai:2018hqu,Christ:2019dxu,Christ:2020hwe}. 
More recently, a lattice calculation near the physical pion mass determined $V(z)=0.87(4.44)$ at $z=0.013(2)$~\cite{RBC:2022ddw}, 
with the sign converted to the convention of Refs.~\cite{DAmbrosio:1998gur,Cirigliano:2011ny}, which we use in the present work and which differs by an overall minus sign from that of Ref.~\cite{RBC:2022ddw}.
The resulting uncertainty is not yet sufficient to resolve the sign of $V(z)$. Variance-reduced estimators for quark loops, including the one-end trick~\cite{Michael:2007vn} (or even-split approach~\cite{Giusti:2019split}), have been introduced to improve the statistical precision of such calculations~\cite{Hodgson:2025split}.

Building on these previous developments, we develop a computational approach for determining $V(z)$ with near-physical pion masses. 
In particular, while previous demonstrations of ultraviolet (UV) finiteness relied on the exact Ward identities satisfied by conserved lattice vector currents, we show that local vector currents, combined with a subtracted Fourier kernel and GIM cancellation, yield UV-finite amplitudes with the same continuum limit as the conserved-current formulation. This substantially simplifies the calculation of nonlocal matrix elements.
We further employ the one-end trick to reduce the stochastic noise from the quark-self-loop contribution. 
To control the exponentially growing term arising from the single-pion intermediate state, we implement four independent subtraction schemes and find consistent results among them.
Using two domain-wall fermion ensembles with near-physical pion masses, we determine $V(z)$ at six values of $z$, spanning both the spacelike and timelike regions, and find
that $V(z)$ is generally negative over the kinematic region studied.

\paragraph{Extraction of $V(z)$}
\label{sec:method} 
The long-distance Minkowski
amplitude is defined as
\begin{equation}
 {\cal A}_{\mu}(q)=\int d^4x\,
 \bra{\pi(p)}T\{\Lweak(x)J_{\mu}(0)\}\ket{K(k)},
 \label{eq:minkowski_amplitude}
\end{equation}
where $k$ and $p$ denote the kaon and pion momenta, respectively, and $q=k-p$ is the
momentum of the virtual-photon.  Electromagnetic gauge invariance, $q^\mu{\cal A}_\mu(q)=0$, constrains the amplitude to take the form:
\begin{equation}
 {\cal A}_{\mu}(q)= -i\frac{G_F}{(4\pi)^2}V(z)
 \left[q^2(k+p)_{\mu}-(M_K^2-M_\pi^2)q_{\mu}\right].
 \label{eq:minkowski_Vz}
\end{equation}
The single form factor $V(z)$ is then obtained from the integrated Minkowski matrix
element.

The $\Delta S=1$ weak Lagrangian is given by
\begin{equation}
 \Lweak=-\frac{G_F}{\sqrt{2}}V_{us}^*V_{ud}
 \sum_{i=1,2}C_i(\mu)\left[Q_i^u(\mu)-Q_i^c(\mu)\right],
 \label{eq:weak_hamiltonian}
\end{equation}
where
\begin{align}
 Q_1^q&=[\bar s\gamma_\nu(1-\gamma_5)d]
       [\bar q\gamma^\nu(1-\gamma_5)q],\nonumber\\
 Q_2^q&=[\bar s\gamma_\nu(1-\gamma_5)q]
       [\bar q\gamma^\nu(1-\gamma_5)d].
 \label{eq:current_current_operators}
\end{align}
We retain only the current-current operators $Q_{1,2}^{q}$ and neglect the contributions from other operators because their matrix elements vanish at tree level and their Wilson coefficients are substantially smaller~\cite{Isidori:2005tv}.
The difference between the up- and charm-quark terms implements the Glashow--Iliopoulos--Maiani (GIM) cancellation.  The four-quark operators $Q_{1,2}^q$
require renormalization, including operator mixing and matching
to the Wilson coefficients.  
We renormalize them nonperturbatively using the RI-SMOM procedure of Ref.~\cite{Christ:2012se}, 
combined with the step-scaling strategy employed in $K\to\pi\pi$ calculations~\cite{Blum:2024Kpipi}. 
Details of the renormalization procedure and the resulting lattice Wilson coefficients are given in the Supplemental Material  (SM)~\cite{SM}.

On the lattice, we compute the Euclidean matrix element using the temporal
component of the vector current,
\ba
&&{\cal A}^E_0=\int_{-T_-}^{T_+} dt_E\,M(t_E,\vec k,\vecp p)
\nn\\
&&M(t_E,\vec k,\vecpp p)=
\nn\\
&&\hspace{0.2cm}\int d^3\vec{x}\, \bra{\pi(\vecpp p)}T\{\Lweak_E(\vec{x},t_E)J_0^E(0)\}\ket{K(\vecp k)},
 \label{eq:matrixelement}
\ea
where the index $E$ denotes ``Euclidean", $T_-$ and $T_+$ denote the limits of the integration in the negative- and positive-$t_E$ directions respectively and $t_E=it$. Our Euclidean conventions are $\gamma_0^E=\gamma^0$ and $\gamma_i^E=-i\gamma^i$, which imply $J_0^E=J^0$ and $\Lweak_E=\Lweak$. The Euclidean matrix element is extracted from the corresponding four-point correlation function with the standard relativistic external-state normalization.

The infinite-time Euclidean integral requires special treatment because the spectral representation contains exponentially growing contributions from intermediate states at negative Euclidean time, $t_E<0$. 
For the quantum numbers of the present matrix element, the two-pion intermediate state does not contribute to the rare-decay amplitude~\cite{Christ:2015aha}. Contributions from the three-pion intermediate state to the exponentially growing behavior are expected to be suppressed by phase space. A more detailed treatment requires phenomenological input for the $K\to\pi\pi\pi$ amplitude and is presented in the SM~\cite{SM}.
The leading exponentially growing term arises from the single-pion intermediate state. We remove this term using four independent subtraction procedures and find consistent results among them, providing a nontrivial check of the subtraction. In each procedure, the physical single-pion contribution to the amplitude is retained or restored. Details of the four subtraction procedures are given in the SM.
At positive Euclidean time, the intermediate states necessarily contain a strange quark and therefore do not give rise to exponentially growing terms. However, when both the kaon in the intermediate state and the pion in the final state carry the same nonzero momentum, the energy gap between them can be small, resulting in slow convergence of the Euclidean-time integral. To address this issue, we reconstruct the large-time contribution using the corresponding matrix element extracted from three-point functions.
After these subtractions and the large-time reconstruction, the Euclidean amplitude ${\cal A}^E_0$ is obtained, from which the Minkowski amplitude is given by ${\cal A}_0=-i{\cal A}^E_0$, and hence the form factor $V(z)$ is determined.

\begin{figure}[!t]
 \centering
 \includegraphics[width=0.98\columnwidth]{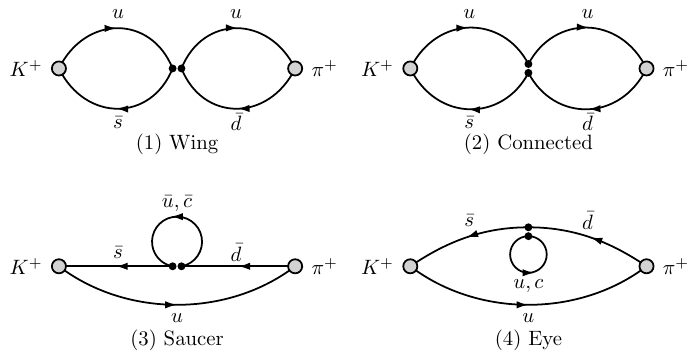}
 \caption{Quark-line contractions of the $K^+\to\pi^+$ weak matrix element: (1) Wing, (2) Connected, (3) Saucer and (4) Eye.}
 \label{fig:contraction-topologies}
\end{figure}

\begin{figure}[t]
 \centering
 \includegraphics[width=0.98\columnwidth]{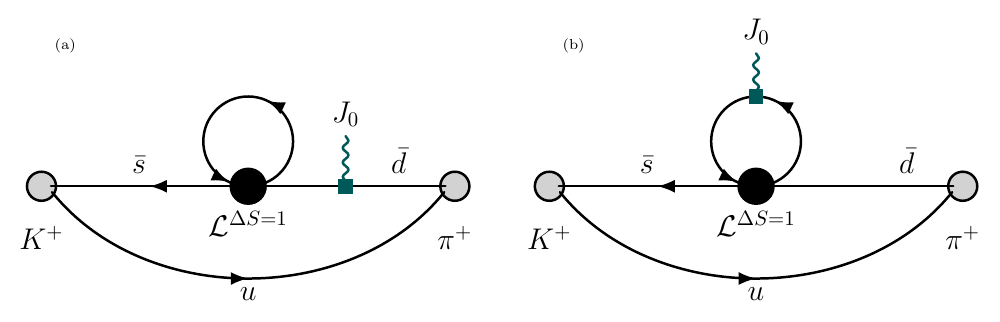}
 \caption{Eye/Saucer contractions with the electromagnetic current inserted (a) outside the loop and (b) on the loop, referred to as the pure-loop and loop-in contractions, respectively. The pure-loop class also includes contractions in which the electromagnetic current is inserted on the external $\bar{s}$ or $u$ quark line; these are not shown explicitly.}
 \label{fig:eye-saucer-contraction}
\end{figure}

The diagrams for $K^+\to\pi^+\ell^+\ell^-$ decays are constructed from the four contraction topologies of the $K^+\to\pi^+$ three-point function, commonly referred to as Eye, Saucer, Wing, and Connected (see Fig.~\ref{fig:contraction-topologies}). The corresponding four-point contractions are obtained by inserting the electromagnetic current on one of the quark lines in Fig.~\ref{fig:contraction-topologies} or by attaching the electromagnetic current to a sea quark.
Particular attention is required for the Eye and Saucer topologies, in which the $q$ and $\bar q$ fields in the four-quark operators $Q_i^q$ are contracted to form a closed quark loop. As illustrated in Fig.~\ref{fig:eye-saucer-contraction}, we distinguish two classes according to the location of the electromagnetic-current insertion. We refer to contractions in which the current is inserted on a quark line outside the $q\bar q$ loop as pure-loop contractions, and those in which the current is inserted on the $q\bar q$ loop as loop-in contractions. These two classes require separate consideration because of their different short-distance behavior.
For the pure-loop contractions, the contributions from lower-dimensional operator mixing do not affect the physical amplitude, while the remaining short-distance contribution is rendered finite by the GIM subtraction. For the loop-in contractions, we reduce the degree of UV divergence arising when the electromagnetic current approaches the weak operator by introducing a subtracted Fourier kernel,
$K(q,t_E,\vec{x})=e^{q^0t_E}e^{-i\vec q\cdot\vec x}-1$.
The corresponding matrix element for the contractions in Fig.~\ref{fig:eye-saucer-contraction}(b) is then evaluated as
\begin{equation}
\begin{aligned}
&\int d^4x\,K(q,-t_E,-\vec{x})\\
&\quad {}\times \left.\bra{\pi}T\{\Lweak_E(0)J_0^E(-t_E,-\vec{x})\}\ket{K}\right|_{\rm loop-in}.
\end{aligned}
\label{eq:subtracted-kernel}
\end{equation} 
Compared to Eq.~(\ref{eq:matrixelement}), we have used spacetime translation invariance to place the weak Lagrangian at the origin and assign the spacetime coordinates $(-t_E,-\vec{x})$ to the vector current.
The subtracted term corresponds to a zero-momentum insertion of the temporal current, so that $\int d^3\vec{x}\,J_0^E(0,-\vecp{x})$ is the charge operator in the continuum theory. Since the loop-in contractions carry zero net charge, this charge-operator insertion vanishes exactly. Thus, the subtraction does not modify the physical amplitude in the continuum theory.

The UV finiteness of the loop-in contractions was previously established using conserved lattice vector currents, whose exact Ward identities reduce the quadratic divergence to a logarithmic one, which is subsequently removed by the GIM subtraction~\cite{Isidori:2005tv}. We extend this result to local vector currents, for which the exact lattice Ward identity is no longer available. Using an OPE analysis, we demonstrate that the subtracted Fourier kernel, together with the GIM subtraction, ensures UV finiteness also for local currents, yielding the same continuum limit as the conserved-current formulation. Details are provided in the SM~\cite{SM}.

Directly estimating the light- and charm-quark loops separately leads to large statistical uncertainties in their GIM-subtracted difference, as observed in previous rare-kaon calculations~\cite{Christ:2016eae,RBC:2022ddw}. To reduce the noise in the pure-loop contractions, we employ the one-end trick~\cite{Michael:2007vn}, also known as the even-split approach~\cite{Giusti:2019split}, following the strategy discussed in Ref.~\cite{Hodgson:2025split}. This method exploits the exact identity
\ba
&&D_u^{-1}(x,x)-D_c^{-1}(x,x)
\nn\\
&=&(m_c-m_u)\sum_yD_u^{-1}(x,y)D_c^{-1}(y,x)
\nn\\
&=&(m_c-m_u)\sum_yD_u^{-1}(x,y)\gamma_5[D_c^{-1}(x,y)]^\dagger\gamma_5.
\label{eq:one-end-identity}
\ea
After generating the propagators $D_{u,c}^{-1}(x,y)$ at all spacetime locations $y$ by using the stochastic volume source and performing the summation at $x$, the light--charm propagator difference is evaluated directly, avoiding the subtraction of two independently noisy loop estimates and thereby substantially reducing the statistical variance.

\paragraph{Numerical analysis} 
\label{sec:numerics} 

We use two near-physical-point domain-wall-fermion ensembles generated by the RBC/UKQCD collaboration~\cite{RBC:2014ntl,RBC:2023xqv}. The 24D ensemble has lattice size $24^3\times64$, $a^{-1}=1.023\,\GeV$, spatial extent $L=4.63\,\fm$, and $M_\pi=143.1(3)\,\MeV$. The 32Dfine ensemble has lattice size $32^3\times64$, $a^{-1}=1.378\,\GeV$, spatial extent $L=4.58\,\fm$, and $M_\pi=142.9(7)\,\MeV$. We use 117 and 73 gauge configurations for the 24D and 32Dfine ensembles, respectively.

Using the random-sparse-field method~\cite{Li:2020hbj,Detmold:2019fbk}, we generate 1024 light- and strange-quark smeared-source propagators per gauge configuration, with source locations randomly selected in spacetime, to construct the kaon and pion interpolating operators at the initial and final states. For the Connected and Wing contractions, we generate an additional 1024 light- and strange-quark point-source propagators per gauge configuration, with randomly selected source positions for the temporal vector current $J_0^E$, while treating $\Lweak_E$ as the sink. This source placement gives substantially smaller statistical uncertainties than the setup with the point source placed at $\Lweak_E$ and $J_0^E$ treated as the sink. The difference is sufficiently large that using the $J_0^E$-centered setup alone yields smaller statistical uncertainties than averaging the two source placements.

The pure-loop and loop-in contractions require charm-quark propagators. For the 24D ensemble, we compute propagators at three different charm-quark masses, corresponding to $\eta_c$ masses of $0.7$, $0.9$, and $1.1\,\GeV$. For the finer 32Dfine ensemble, we use one additional charm-quark mass corresponding to $M_{\eta_c}=1.3\,\GeV$. The loop-in contractions have substantially better signal-to-noise ratios than the pure-loop contractions. We use the same source strategy as for the Connected and Wing contractions, with the point source placed at $J_0^E$. For each charm-quark mass, we generate 512 point-source propagators per gauge configuration. For the statistically most challenging pure-loop contractions, we instead generate 256 $U(1)$ stochastic-volume-source propagators per gauge configuration for each charm-quark mass and employ the error-reduction strategy described above. This substantially improves the statistical precision of the pure-loop contractions, reducing the statistical uncertainties at $t_E=0$ by factors of approximately $5$--$16$, depending on the charm-quark mass and ensemble. A detailed comparison with and without error reduction is presented in the SM~\cite{SM}.

Smeared kaon and pion interpolating operators are used for all the contractions, which provide substantially better overlap with the ground states than point interpolating operators. For both the kaon and pion, the temporal separations from the nearest operators $\Lweak_E$ or $J_0^E$ are fixed at $6a\simeq1.16\,\fm$ on 24D and $8a\simeq1.15\,\fm$ on the 32Dfine ensemble, with these separations chosen to achieve adequate ground-state dominance. Local vector currents are used for $J_0^E$, with the corresponding renormalization factors $Z_V$ taken from Ref.~\cite{Feng:2021zek}.

\begin{figure}[t]
 \centering
 \includegraphics[width=\columnwidth]{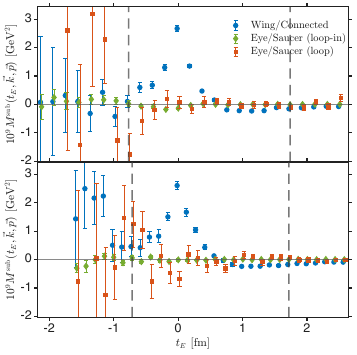}
 \caption{Time dependence of the matrix element after one-pion subtraction using Method 3 for $K(0)\to\pi(p)$ at $|\vecp p|=2\pi/L$ on the 24D and 32Dfine ensembles.}
\label{fig:paired-time}
\end{figure}

The matrix elements after one-pion subtraction are presented in Fig.~\ref{fig:paired-time}. As an example, 
we consider the kinematics $\vec{k}=\vec{0}$ and $|\vecp{p}|=2\pi/L$. The results are averaged over the six momentum directions related by cubic rotational symmetry,
 namely $(\pm1,0,0)$, $(0,\pm1,0)$, and $(0,0,\pm1)$.
 The figure shows the $t_E$ dependence of the subtracted matrix element $M^{\rm sub}(t_E,\vec{k},\vecp{p})$, where the superscript ``sub'' indicates that the one-pion subtraction has been performed.
For the one-pion subtraction, we employ the four independent methods described in the SM~\cite{SM}. The results shown in Fig.~\ref{fig:paired-time} use Method 3. 
Among the contraction classes shown in Fig.~\ref{fig:paired-time}, the Wing/Connected and pure-loop contributions require one-pion subtraction, whereas the loop-in contribution does not. The Wing/Connected contribution dominates the signal, while the two Eye/Saucer contributions are small, with the pure-loop contribution carrying the largest statistical uncertainty.

For a direct comparison between the 24D and 32Dfine ensembles, the charm-quark mass in Fig.~\ref{fig:paired-time} is chosen such that $M_{\eta_c}=1.1\,\GeV$ for both ensembles. This value is still substantially below the physical charm-quark mass. The dependence on the charm-quark mass, including values up to $M_{\eta_c}=1.3\,\GeV$, and the extrapolation toward the physical charm-quark mass are investigated in the SM~\cite{SM}. 
The two vertical dashed lines indicate the values of $T_-$ and $T_+$ used in the final integration. At negative Euclidean time, we employ a modified weak Lagrangian, defined in the SM~\cite{SM}, which redistributes the single-pion contribution among other intermediate states, thereby eliminating the unphysical exponentially growing term without changing the physical amplitude. The remaining excited-state contributions decay rapidly, and the matrix element is already saturated by the value of $T_-$ shown in Fig.~\ref{fig:paired-time}. 
At positive Euclidean time, the energy gap between the kaon intermediate state and the pion final state, $E_K(\vecp p)-E_\pi(\vecp p)$, is small, resulting in a slow convergence of the Euclidean-time integral. In particular, the Wing/Connected contribution changes sign from positive to negative as $t_E$ increases. We confirm this long-distance behavior through a reconstruction based on the corresponding matrix element determined from three-point functions. With this reconstruction, the upper integration limit $T_+$ can be chosen at the value indicated in Fig.~\ref{fig:paired-time}, rather than extending the integration to the roughly $3\,\fm$ range otherwise required for convergence. The dependence on $T_-$ and $T_+$ has been systematically checked and is presented in the SM~\cite{SM}.

We obtain the form factor $V(z)$ by performing the Euclidean-time integral of the subtracted matrix element and combining the contributions from all contraction classes. The resulting lattice data for $V(z)$ are then fitted using polynomial parameterizations of increasing order. To assess the sensitivity to the assumed momentum dependence, we consider fits up to quadratic order in $z$, both including and excluding the phenomenological contribution $V^{\pi\pi}(z)$. 
We extrapolate the lattice results to the physical charm-quark mass, $M_{\eta_c}=M_{\eta_c}^{\rm phys}$, using a logarithmic ansatz motivated by the GIM-subtracted short-distance behavior and consistent with the observed charm-mass dependence. The resulting fit parameters, values of $V(0)$, and $\chi^2/\mathrm{dof}$ are summarized in the SM~\cite{SM}.
The negative sign of $V(0)$ is stable against variations in the fit order, the inclusion or exclusion of $V^{\pi\pi}(z)$, and the extrapolation to the physical charm-quark mass. The charm-mass extrapolation leads to a moderate shift in the central value and a more substantial increase in the statistical uncertainty, but $V(0)$ remains negative for all fit ans\"atze on both ensembles.

\begin{figure}[t]
 \centering
 \includegraphics[width=\columnwidth]{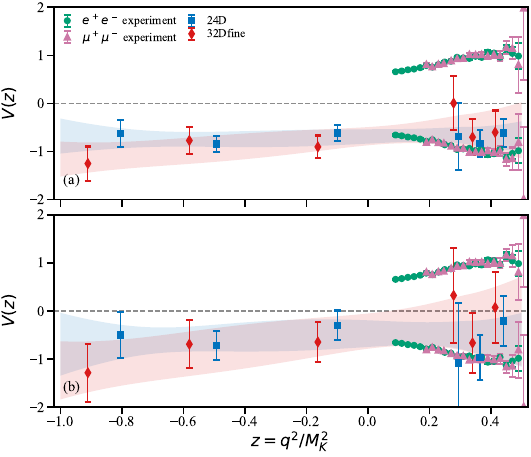}
 \caption{Lattice results for $V(z)$ using Method 3 for the spacelike points and Method 2 for the timelike points, compared with the positive and negative branches of the experimental data. The bands indicate the uncertainties from the $a_+ + b_+z + c_+z^2 + V^{\pi\pi}(z)$ fits. The upper panel shows the lattice results at $M_{\eta_c}=1.1\,\GeV$, while the lower panel shows the results extrapolated to $M_{\eta_c}=M_{\eta_c}^{\rm phys}$.}
 \label{fig:paired-v-experiment}
\end{figure}

Fig.~\ref{fig:paired-v-experiment} compares our lattice results for $V(z)$ and the fits using $a_+ + b_+z + c_+z^2 + V^{\pi\pi}(z)$ with the experimental data. Here $V^{\pi\pi}(z)$ is a known function of $z$, so that $a_+$, $b_+$, and $c_+$ are the only fit parameters.
Both signs of $V(z)$ allowed by the experimental measurements are shown. Of the six momentum modes considered, the three with $\vec{k}=\vec{0}$ and $\vec{p}\neq\vec{0}$ lie in the spacelike region and are determined more precisely, whereas the three with $\vec{k}\neq\vec{0}$ and $\vec{p}=\vec{0}$ lie in the timelike region. The experimental data, in contrast, are restricted to the timelike region. Lattice QCD and experimental measurements therefore provide complementary information on the momentum dependence of $V(z)$.
With only two lattice spacings and the present statistical precision, we do not attempt a continuum extrapolation. Taking $\Lambda_{\rm QCD}\sim500\,\mathrm{MeV}$, a naive estimate of $O((\Lambda_{\rm QCD}a)^2)$ discretization effects gives corrections of order 10\%--20\%. The consistency between the 24D and 32Dfine results provides a useful check, but a controlled determination of discretization effects will require additional lattice spacings.
The experimental measurements determine $|V(z)|$ much more precisely in the timelike region, but do not determine its sign. The present lattice results have substantially larger statistical uncertainties, yet provide direct information on the sign of $V(z)$ and favor the negative branch. The uncertainty bands from the lattice fits encompass the negative branch of the experimental data over the kinematic range shown.

\paragraph{Conclusion}

We have presented a lattice QCD calculation of the form factor $V(z)$ governing the rare decay $K^+\to\pi^+\ell^+\ell^-$. Using two near-physical-point domain-wall-fermion ensembles and an improved computational strategy, we obtain statistically significant signals for $V(z)$ at six momentum transfers spanning both the spacelike and timelike regions. Treating the statistical uncertainties on the two ensembles as independent, we combine the fitted values of $V(0)$ using inverse-variance weighting. The resulting significance for a negative $V(0)$ ranges from $5.8\sigma$ to $7.9\sigma$ at $M_{\eta_c}=1.1\,\mathrm{GeV}$, depending on the fit ansatz, and is reduced to $2.0\sigma$ - $3.4\sigma$ after extrapolation to  the physical charm quark mass.
In the timelike region, the lattice results are also consistent with the negative branch extracted from experiment, but disfavor the positive branch.
These results provide direct first-principles information on a sign that is not determined by the experimental decay rate alone.

Lattice studies of this decay began more than a decade ago, but a fully controlled calculation remains challenging because simultaneously reaching the physical pion and charm-quark masses requires both a sufficiently large spatial volume and a sufficiently fine lattice spacing. In the present calculation, the pion mass is near its physical value, while the charm-quark mass remains below the physical point. The present uncertainties also remain substantially larger than those from experiment. Further improvements will require finer lattice spacings to accommodate heavier charm quarks and enable a controlled continuum extrapolation, together with increased statistics and improved noise-reduction techniques for the pure-loop contractions. These developments will help reduce both statistical and systematic uncertainties in future determinations of $V(z)$.

Beyond these improvements, the lattice approach offers information complementary to experiment: it is not restricted to the physical timelike region accessible in the decay, and the spacelike results provide additional constraints on the momentum dependence of $V(z)$. More broadly, the computational strategy developed here provides a framework for first-principles studies of other long-distance rare weak processes,  including rare kaon decays such as $K_S\to\pi^0\ell^+\ell^-$ and $K\to\pi\nu\bar{\nu}$, as well as rare hyperon decays. Together with increased precision and improved control of systematic effects, these developments open a path toward fully controlled lattice studies of a broader class of rare hadronic decays.

\begin{acknowledgements}
We gratefully acknowledge many helpful discussions with our colleagues from the RBC-UKQCD Collaborations.  X.F., B.L.L., C.L., and Z.L.Z. were supported in part by the National Natural Science Foundation of China under Grants No.~12125501, No.~12550007,
No.~125B2104], No.~12293060, and No.~12293063.  L.C.J. acknowledges support by the U.S.
Department of Energy Office of Science Early Career Award DE-SC0021147, DOE
grants DE-SC0010339 and DE-SC0026314. C.T.S. was supported in part by STFC (UK) grant ST/X000583/1.  The research reported in this work was
carried out using the computing facilities at the Chinese National Supercomputer
Center in Tianjin.  It also made use of computing and long-term storage
facilities of the USQCD Collaboration, which are funded by the Office of
Science of the U.S. Department of Energy.
\end{acknowledgements}

\bibliography{ref}

\clearpage
\onecolumngrid
\clearpage
\twocolumngrid
\clearpage
\setcounter{page}{1}
\renewcommand{\thepage}{Supplementary Information -- S\arabic{page}}
\setcounter{equation}{0}
\renewcommand{\theequation}{S\,\arabic{equation}}
\renewcommand{\theHequation}{S\arabic{equation}}
\setcounter{figure}{0}
\renewcommand{\thefigure}{S\,\arabic{figure}}
\renewcommand{\theHfigure}{S\arabic{figure}}
\setcounter{table}{0}
\renewcommand{\thetable}{S\,\Roman{table}}

\section{Supplementary Information}

\subsection{Contribution from the Wing and Connected contractions}

\begin{figure}[htb]
 \centering
 \includegraphics[width=\columnwidth]{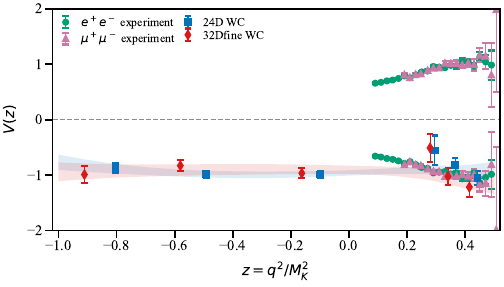}
 \caption{Form factor $V(z)$ obtained using only the Wing and Connected contractions, with the loop-in and pure-loop contributions omitted. The substantially reduced statistical uncertainties reveal a clear preference for the negative branch. This comparison illustrates the dominant contribution of the Wing and Connected contractions to the signal and the significant statistical noise introduced by the loop contractions.}
 \label{fig:figure_4_Vz_WC}
\end{figure}

To assess the impact of the loop-in and pure-loop contractions on the determination of the form factor, we repeat the analysis retaining only the Wing and Connected contractions. As shown in Fig.~\ref{fig:figure_4_Vz_WC}, the resulting form factor strongly favors the negative branch, with substantially smaller statistical uncertainties than those of the full calculation. This confirms that the Wing and Connected contractions dominate the signal, whereas the loop-in and pure-loop contractions contribute significantly to the statistical uncertainty. Although the Wing-and-Connected-only result does not represent the complete physical amplitude, this comparison provides a useful illustration of the statistical challenges associated with the disconnected contractions.

\subsection{Error reduction for pure-loop correlators}

\begin{figure}[htb]
 \centering
 \includegraphics[width=\columnwidth]{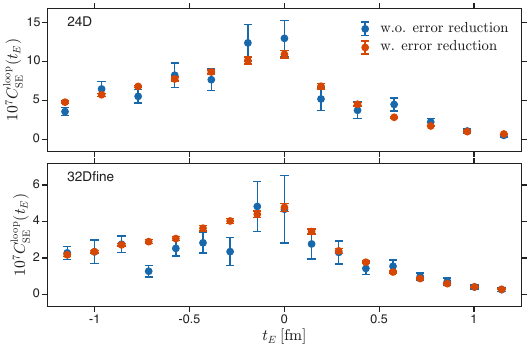}
 \caption{Pure-loop correlators with and without error reduction on 24D (upper) and 32Dfine (lower), at $M_{\eta_c}=1.1$ and $1.3\,\GeV$, respectively.}
 \label{fig:se-even-split-comparison}
\end{figure}

We employ the error-reduction strategy known as one-end trick or even-split approach for pure-loop correlators. The comparison with and without error reduction is shown in Fig.~\ref{fig:se-even-split-comparison}. At $t_E=0$, the statistical error is reduced by factors of about 7, 6 and 5
across the three charm-quark masses on the 24D ensemble, and
16, 13, 10 and 9 
across the four charm-quark masses on the 32Dfine ensemble, in order of increasing charm-quark mass.

\subsection{Comparison of one-pion subtraction schemes}

At negative Euclidean time, the one-pion intermediate state produces an
exponentially growing contribution. 
We employ four subtraction procedures, the first two of which were developed in previous lattice studies of rare kaon decays~\cite{Christ:2015aha,Christ:2016eae,RBC:2022ddw}.

\paragraph{Method 1.} 
The most direct approach is to reconstruct the large-$T_-$ contribution using matrix elements extracted from three-point functions. The single-pion contribution to the time-integrated four-point function is proportional to
\ba
&&\langle \pi(\vecp{p})|J_0^E(0)|\pi(\vecp{k})\rangle
\langle \pi(\vecp{k})|\Lweak_E(0)|K(\vecp{k})\rangle
\nn\\
&&\hspace{2.5cm}\times
\frac{1-e^{[E_K(\vec{k})-E_\pi(\vec{k})]T_-}}
{E_K(\vec{k})-E_\pi(\vec{k})}.
\ea
We replace the finite-$T_-$ factor by its infinite-time limit,
$\frac{1}{E_K(\vec{k})-E_\pi(\vec{k})}$,
using the corresponding matrix elements determined from three-point functions. This removes the dependence on the finite temporal extent $T_-$.
In practice we subtract this single-pion contribution slice by slice,
$M^{\rm sub}(t_E)=M(t_E)-B\,e^{-[E_K(\vec{k})-E_\pi(\vec{k})]t_E}$ for $t_E<0$, so that the residual is exponentially suppressed and integrable, and then add back the reconstructed $B/[E_K(\vec{k})-E_\pi(\vec{k})]$, with
$B=\langle \pi(\vecp{p})|J_0^E|\pi(\vecp{k})\rangle\langle \pi(\vecp{k})|\Lweak_E|K(\vecp{k})\rangle$
taken from the fitted three-point matrix elements. A disadvantage of this approach is that the matrix elements entering the reconstruction are obtained independently from three-point functions and are therefore not optimally correlated with the four-point data, limiting the cancellation of statistical and systematic uncertainties.

\paragraph{Method 2.}
A second approach exploits the fact that the scalar density $\bar{s}d$ can be related to the divergence of the vector current. Adding a term proportional to $\bar{s}d$ to the weak Lagrangian therefore does not alter the physical rare-decay amplitude. We define
\be
{\Lweak_E}'(x)=
\Lweak_E(x)-c\,\bar{s}d(x),
\ee
where
\be
c=
\frac{\langle\pi(\vecp{k})|\Lweak_E(0)|K(\vecp{k})\rangle}
{\langle\pi(\vecp{k})|\bar{s}d(0)|K(\vecp{k})\rangle}.
\ee
The coefficient $c$ is determined from the ratio of the corresponding three-point functions, retaining their statistical correlation.
With this choice, the single-pion matrix element of the modified weak Lagrangian vanishes, 
thereby removing the leading exponentially growing contribution.
Although this modification changes the contributions of individual intermediate states, their sum, and hence the physical decay amplitude, remains unchanged.

\paragraph{Method 3.}
A third approach determines the subtraction coefficient directly from the four-point functions by exploiting charge conservation. 
We choose the kinematics $\vec p=\vec k$, such that the vector current carries zero momentum and $J_0^E$ reduces to the charge operator.
We define
\be
c(t_E)=
\frac{M(t_E,\vec{k},\vecp{k})}
{M_{\rm SD}(t_E,\vec{k},\vecp{k})},
\ee
where $M_{\rm SD}(t_E,\vec{k},\vecp{p})$ is obtained from $M(t_E,\vec{k},\vecp{p})$ by replacing $\Lweak_E$ with $\bar{s}d$. For a conserved vector current, $c(t_E)$ is independent of $t_E$. With the local vector current used in our calculation, residual lattice artifacts can induce a mild $t_E$ dependence at small $t_E$, consistent with the enhanced lattice artifacts expected in the short-distance region.
The subtraction is then implemented as
\be
M^{\rm sub}(t_E,\vec{k},\vec{p})
= 
M(t_E,\vec{k},\vecp{p})
-c(t_E)M_{\rm SD}(t_E,\vecp{k},\vecp{p}).
\ee

An advantage of this construction is that the subtraction coefficient and the original matrix element are determined from strongly correlated measurements, leading to a substantial correlation between $M(t_E,\vec{k},\vecp{p})$ and the subtraction term $c(t_E)M_{\rm SD}(t_E,\vec{k},\vecp{p})$. This correlation enhances the cancellation of statistical fluctuations and common systematic effects in the subtracted matrix element. For the spacelike kinematics, $\vec{k}=\vec{0}$ and $\vec{p}\neq\vec{0}$, we therefore use Method 3 for our final results. For the timelike kinematics, however, $\vec{k}\neq\vec{0}$, and the determination of $c(t_E)$ involves the denominator $M_{\rm SD}(t_E,\vec{k},\vecp{k})$ at nonzero momentum, which is statistically noisier. We therefore use Method 2 for our final results in the timelike region.

\paragraph{Method 4.}
The fourth approach avoids the computation of the scalar-density four-point function. We determine the subtraction coefficient from
\[
c(\vec{k},\vecp{p})=
\lim_{t_E\to-\infty}
\frac{M(t_E,\vec k,\vecp p)}
{M(t_E,\vec k,\vecp k)},
\]
and define
\[
M^{\rm sub}(t_E,\vec k,\vec p)
=
M(t_E,\vec k,\vec p)
-c(\vec{k},\vec{p})\,M(t_E,\vec k,\vec k).
\]
The key observation is that $M(t_E,\vec k,\vec k)$ does not contribute to the physical decay amplitude, as follows from the Lorentz structure in Eq.~\eqref{eq:minkowski_Vz}. Nevertheless, it contains the same leading exponentially growing term arising from the single-pion intermediate state. Subtracting this term therefore removes the exponential growth without affecting the physical amplitude.

\begin{figure}[htb]
 \centering
 \includegraphics[width=\columnwidth]{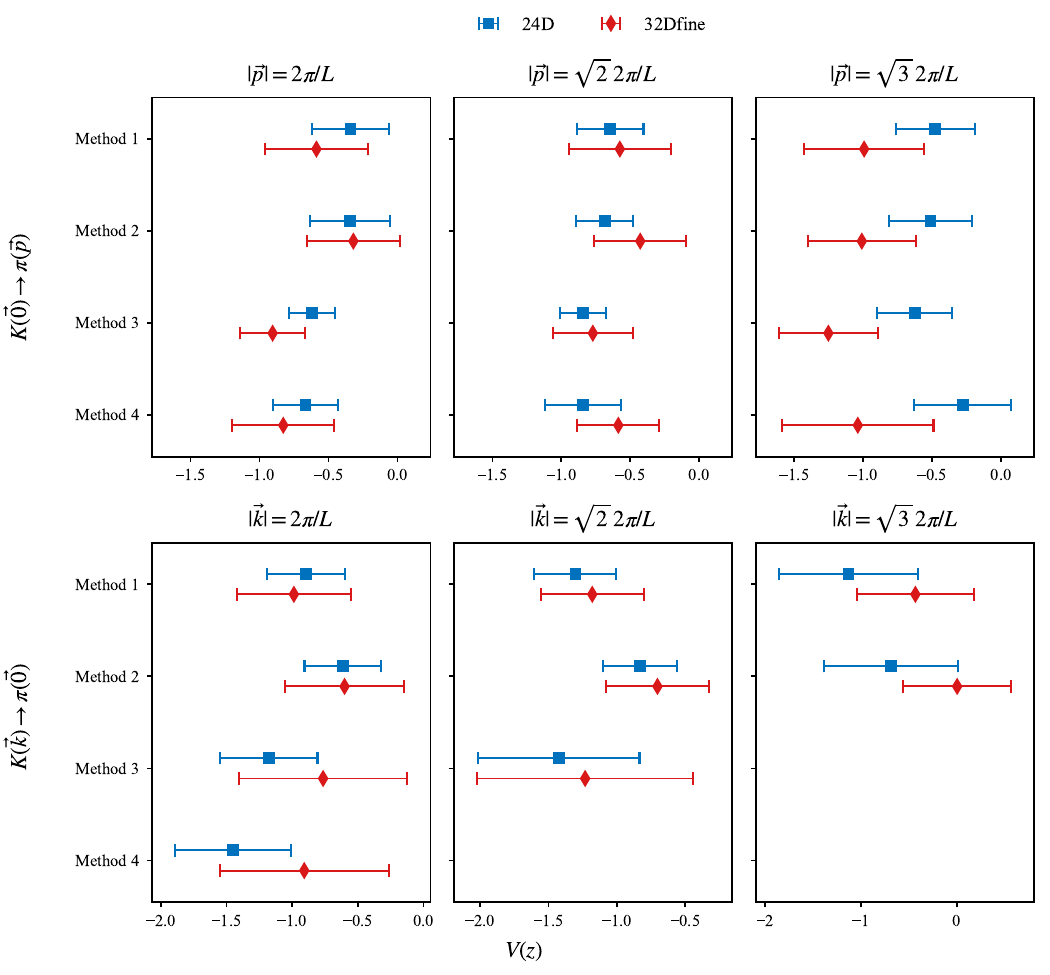}
 \caption{Comparison of one-pion subtraction methods for $K(0)\to\pi(\vecp{p})$ (upper row) and $K(\vec{k})\to\pi(\vecp{0})$ (lower row). Columns show $|\vecp p|$ or $|\vec{k}|=(1,\sqrt{2},\sqrt{3})\,2\pi/L$.  For the timelike momentum modes, the denominator entering the ratio that determines the subtraction coefficient becomes increasingly noisy, making the ratio unstable at higher momenta. Method 3 is therefore shown only for the two lowest momenta, and Method 4 only for the lowest momentum.}
 \label{fig:onepion-method-comparison}
\end{figure}

We compare the four subtraction procedures described above using the corresponding three-point and four-point correlation functions. Fig.~\ref{fig:onepion-method-comparison} compares $V(z)$ for $K(\vecp{0})\to\pi(\vecp{p})$ in the upper row and $K(\vecp{k})\to\pi(\vecp{0})$ in the lower row. The three columns correspond to $|\vecp p|$ or $|\vecp{k}|=(1,\sqrt2,\sqrt3)\,2\pi/L$; the vertical positions distinguish the subtraction methods, and the two marker colours distinguish the ensembles.
The various methods give consistent results within statistical uncertainties, despite their different implementations and sources of statistical correlations. This agreement provides a nontrivial check of the treatment of the exponentially growing single-pion contribution.

\subsection{Dependence on $T_-$ and $T_+$}

\begin{figure}[htb]
 \centering
 \includegraphics[width=\columnwidth]{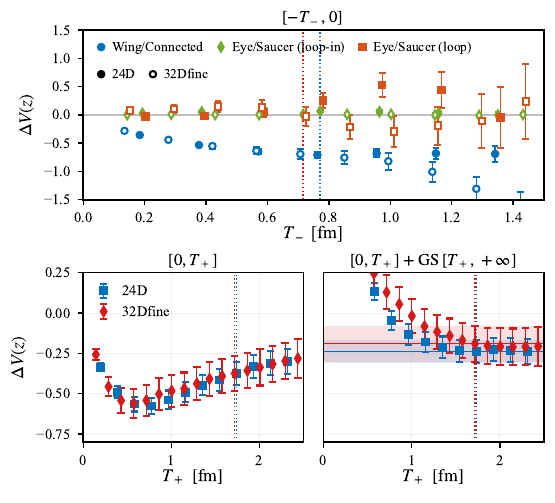}
 \caption{Negative- and positive-time contributions $\Delta V(z)$ obtained using Method 3 for $K(\vecp{0})\to\pi(\vecp{p})$ at $|\vecp p|=2\pi/L$. 
 The upper panel shows the negative-time contribution integrated over $[-T_-,0]$ as a function of $T_-$, with the Wing \& Connected, loop-in, and pure-loop contributions shown separately. The lower panels show the sum of all three contributions: the lower-left panel shows the positive-time contribution integrated over $[0,T_+]$, while the lower-right panel includes in addition the ground-state reconstruction of the long-distance tail from $T_+$ to infinity. Blue and red denote the 24D and 32Dfine ensembles, respectively. Dotted lines mark the adopted integration limits, and horizontal bands indicate the corresponding values and statistical uncertainties.}
 \label{fig:tcut-dependence}
\end{figure}

The form factor $V(z)$ is obtained by integrating the matrix element over both the negative- and positive-$t_E$ regions, from $-T_-$ to $T_+$. As discussed in the main text, the modification of the weak Lagrangian eliminates the one-pion contribution to the spectral decomposition at negative Euclidean time, leaving only the rapidly decaying excited-state contributions. At positive Euclidean time, the long-distance contribution is reconstructed from the corresponding matrix element obtained from three-point functions. The integrated amplitude is therefore expected to become insensitive to $T_-$ and $T_+$ once the remaining excited-state contributions are sufficiently suppressed and the long-distance tail is reliably reconstructed.

To examine the dependence on the two integration limits separately, we consider the contributions integrated over $[-T_-,0]$ and $[0,T_+]$, respectively, and denote them by $\Delta V(z)$ to distinguish them from the full form factor $V(z)$. 
As shown in the upper panel of Fig.~\ref{fig:tcut-dependence}, the negative-time contribution is dominated by the Wing \& Connected diagrams, while the loop-in and pure-loop contributions are much smaller in magnitude but dominate the statistical uncertainty. For the most precise contribution, the Wing \& Connected diagrams on the 24D ensemble, a clear plateau is observed around the adopted $T_-$, indicating that the excited-state contributions are sufficiently suppressed. The 32Dfine data exhibit an anomalous downward shift at larger $T_-$. To investigate its origin, we examine the unintegrated matrix element shown in Fig.~\ref{fig:paired-time}. At $t_E=-0.86$ and $-1.00$ fm, the matrix element is consistent with zero within $0.81\sigma$ and $0.90\sigma$, respectively. However, at larger separations of $t_E=-1.15$, $-1.29$, and $-1.43$ fm, it exhibits deviations of $3.1\sigma$, $2.3\sigma$, and $1.8\sigma$, respectively. This nonmonotonic behavior is suggestive of statistical fluctuations rather than a smoothly decaying long-distance contribution. Given the small magnitudes of the loop-in and pure-loop contributions, we do not expect their long-distance tails to introduce substantial truncation effects. To estimate the residual truncation uncertainty, we compare the 24D Wing \& Connected results at $T_-\simeq 0.8$ fm and $1.4$ fm. The difference is negligible compared with the statistical uncertainty of the total contribution at the adopted $T_-$, suggesting that the associated truncation effects are small at the present level of precision.

The lower panels of Fig.~\ref{fig:tcut-dependence} illustrate the effect of the long-distance reconstruction. Without reconstruction, the integrated contribution exhibits a clear dependence on $T_+$, whereas including the reconstructed long-distance tail leads to a stable plateau over the range used in our analysis. This demonstrates that the reconstruction effectively removes the dependence on the positive-time integration cutoff. We therefore adopt the values of $T_-$ and $T_+$ specified in the main text for our final results.

\subsection{Renormalization of the four-quark operators}
\label{sec:sm-nf4-renormalization}

For the 24D and 32Dfine ensembles, we combine the nonperturbative renormalization procedure of Ref.~\cite{Christ:2012se} 
with the step-scaling strategy used in $K\to\pi\pi$ calculations~\cite{Blum:2024Kpipi}. 
We consider the GIM-subtracted four-quark operators
$\mathbf Q=(Q_1^u-Q_1^c,Q_2^u-Q_2^c)^T$ in the
$\mathrm{RI\! -\! SMOM}(\gamma_\mu,\slashed{q})$ scheme.
With the Wilson coefficients written as a row vector, the renormalization and matching relations are
\begin{align}
 \mathbf Q^{\rm RI}(\mu_l)&=Z^{\rm RI}(\mu_l,a)\mathbf Q^{\rm lat}(a),\\
 \mathbf Q^{\overline{\rm MS}}(\mu_h)&=R(\mu_h)\Sigma(\mu_h,\mu_l)
 Z^{\rm RI}(\mu_l,a)\mathbf Q^{\rm lat}(a),\\
 \mathbf C^{\rm lat}(a)&=\mathbf C^{\overline{\rm MS}}(\mu_h)
 R(\mu_h)\Sigma(\mu_h,\mu_l)Z^{\rm RI}(\mu_l,a).
 \label{eq:sm-nf4-master}
\end{align}
Here $R(\mu_h)$ converts the operators from RI-SMOM to $\overline{\rm MS}$, and all matrices act on the ordered $(Q_1,Q_2)$ basis. We quote the renormalization matrices to three significant digits; their statistical uncertainties are negligible at this precision and are therefore omitted.

\paragraph{Low-scale RI-SMOM matrices.}
We consider two low-scale routes, denoted by $\mu_{l1}\simeq1.3\,\mathrm{GeV}$ and $\mu_{l2}\simeq1.5\,\mathrm{GeV}$. 
The RI-SMOM scales are set by the lattice momenta used in the NPR calculation and therefore differ slightly between ensembles when expressed in physical units. 
The corresponding scales on the (24D, 32Dfine) ensembles are $(1.31,1.32)\,\mathrm{GeV}$ for the $\mu_{l1}$ route and $(1.52,1.52)\,\mathrm{GeV}$ for the $\mu_{l2}$ route. The corresponding values are
\begin{equation}
 Z^{\rm RI}_{24D}(\mu_{l1})=\begin{pmatrix}0.672 & -0.0837\\-0.0837 & 0.672\end{pmatrix},
\end{equation}
\begin{equation}
 Z^{\rm RI}_{32Dfine}(\mu_{l1})=\begin{pmatrix}0.577 & -0.0445\\-0.0445 & 0.577\end{pmatrix},
\end{equation}
\begin{equation}
 Z^{\rm RI}_{24D}(\mu_{l2})=\begin{pmatrix}0.668 & -0.0935\\-0.0935 & 0.668\end{pmatrix},
\end{equation}
\begin{equation}
 Z^{\rm RI}_{32Dfine}(\mu_{l2})=\begin{pmatrix}0.573 & -0.0550\\-0.0550 & 0.573\end{pmatrix}.
\end{equation}

\paragraph{Step scaling on 32IfineH.}
The step-scaling matrices determined on the 32IfineH ensemble ($a^{-1}=1.785(5)$ GeV) are defined by
\begin{equation}
 \Sigma(\mu_h,\mu_l)=
 Z^{\rm RI}_{\rm fine}(\mu_h)
 [Z^{\rm RI}_{\rm fine}(\mu_l)]^{-1},
\end{equation}
with the lattice high scale $\mu_h=2.14\,\mathrm{GeV}$.
For the route through $\mu_{l1}$,
\begin{equation}
 \Sigma(\mu_h,\mu_{l1})=\begin{pmatrix}0.945 & -0.0795\\-0.0795 & 0.945\end{pmatrix}.
\end{equation}
and for the route through $\mu_{l2}$,
\begin{equation}
 \Sigma(\mu_h,\mu_{l2})=\begin{pmatrix}1.04 & -0.0514\\-0.0514 & 1.04\end{pmatrix}.
\end{equation}

We use these step-scaling matrices without a continuum extrapolation. The dependence on the choice of low-scale route is propagated to the final lattice Wilson coefficients 
as an estimate of the renormalization systematic uncertainty.

For the route through $\mu_{l1}$, the resulting matrices $Z^{\rm RI}(\mu_h,a)=\Sigma(\mu_h,\mu_l)Z^{\rm RI}(\mu_l,a)$  are
\begin{equation}
 Z^{\rm RI}_{24D}(\mu_h)=\begin{pmatrix}0.641 & -0.133\\-0.133 & 0.641\end{pmatrix}.
\end{equation}
\begin{equation}
 Z^{\rm RI}_{32Dfine}(\mu_h)=\begin{pmatrix}0.548 & -0.0879\\-0.0879 & 0.548\end{pmatrix},
\end{equation}
while those for the route through $\mu_{l2}$ are
\begin{equation}
 Z^{\rm RI}_{24D}(\mu_h)=\begin{pmatrix}0.701 & -0.132\\-0.132 & 0.701\end{pmatrix}.
\end{equation}
\begin{equation}
 Z^{\rm RI}_{32Dfine}(\mu_h)=\begin{pmatrix}0.600 & -0.0868\\-0.0868 & 0.600\end{pmatrix}.
\end{equation}

\paragraph{Conversion to $\overline{\rm MS}$.}
For $R(\mu_h)$, we use the one-loop RI-SMOM-to-$\overline{\rm MS}$
conversion given in Eq.~(26) of Ref.~\cite{Christ:2012se}.
The perturbative matching inputs are evaluated at $2.15\,\mathrm{GeV}$, 
whose difference from the lattice high scale $\mu_h=2.14\,\mathrm{GeV}$ is negligible at the precision considered here. 
The resulting lattice Wilson coefficients are listed in Table~\ref{tab:sm-nf4-coefficients}.

\begin{table}[htb]
\centering
\setlength{\tabcolsep}{5pt}
\begin{ruledtabular}
\fontsize{8}{9.5}\selectfont
\begin{tabular}{c r r}
Ensemble & $C_1^{\rm lat}$ & $C_2^{\rm lat}$ \\
\hline
\multicolumn{3}{c}{$\mathrm{RI\! -\! SMOM}(\gamma_\mu,\slashed{q})$} \\
24D & $-0.313(15)$ & $0.716(63)$ \\
32Dfine & $-0.241(13)$ & $0.606(55)$ \\
\hline
\multicolumn{3}{c}{$\mathrm{RI\! -\! SMOM}(\gamma_\mu,\gamma_\mu)$} \\
24D & $-0.293(11)$ & $0.6699(16)$ \\
32Dfine & $-0.2279(63)$ & $0.5731(40)$ \\
\hline
\multicolumn{3}{c}{$\mathrm{RI\! -\! SMOM}(\slashed{q},\slashed{q})$} \\
24D & $-0.3278(46)$ & $0.73(13)$ \\
32Dfine & $-0.2513(12)$ & $0.61(10)$ \\
\end{tabular}
\end{ruledtabular}
\caption{Lattice Wilson coefficients obtained using three intermediate RI-SMOM schemes. Central values are obtained using the  $\mu_{l1}$ route, while the values in parentheses indicate the absolute shifts when the $\mu_{l2}$ route is used instead.}
\label{tab:sm-nf4-coefficients}
\end{table}

In addition to the \mbox{$\text{RI-SMOM}(\gamma_\mu,\slashed{q})$} scheme adopted in our central analysis, we also calculate the renormalization factors in the \mbox{$\text{RI-SMOM}(\gamma_\mu,\gamma_\mu)$} and \mbox{$\text{RI-SMOM}(\slashed{q},\slashed{q})$} schemes, as summarized in Table~\ref{tab:sm-nf4-coefficients}. The differences among these schemes provide an estimate of the systematic uncertainty associated with the NPR procedure. Similar levels of scheme dependence have been observed in previous $K\to\pi\pi$ calculations~\cite{Blum:2024Kpipi}. At the present precision, these variations do not affect our conclusion regarding the negative sign of $V(z)$.

\subsection{UV behavior of the pure-loop contraction}

For the pure-loop contraction, we examine the short-distance behavior of the weak Lagrangian through the possible mixing of the left-handed four-quark operators $Q_i^q$ ($i=1,2$) with lower-dimensional operators, taking into account the symmetries relevant to the $K\to\pi\ell^+\ell^-$ amplitude.

For the $K\to\pi$ matrix element, a class of operators, including the pseudoscalar density $\bar{s}\gamma_5d$, do not contribute because of parity. The lowest-dimensional operator that can contribute is therefore the scalar density $\bar{s}d$. Since this operator flips chirality, its mixing with the left-handed weak operators vanishes in the chiral limit. At leading order in the external-flavor quark masses, the mixing must consequently be proportional to $(c_1 m_s+c_2 m_d)\bar{s}d$.

The coefficients $c_1$ and $c_2$ are further constrained by CPS symmetry, defined as charge conjugation and parity followed by the interchange $s\leftrightarrow d$, including the corresponding quark masses~\cite{Bernard:1985wf}. The weak operators $Q_i^q$ are CPS even, as is the scalar density $\bar{s}d$. Under CPS, the mass-dependent coefficient transforms as $c_1 m_s+c_2 m_d\to c_1 m_d+c_2 m_s$. CPS invariance therefore requires $c_1=c_2$, leaving the combination $(m_s+m_d)\bar{s}d$. The chromomagnetic operator, which also flips chirality, requires an additional quark-mass insertion and can thus be included among the dimension-six operators. The resulting mixing structure is
\begin{equation}
 \begin{aligned}
 \left.Q_i^q\right|_{\rm loop}
 &\underset{\rm mixing}{\longrightarrow}
 C_{i,S}^q(\Lambda,m_q)\,(m_s+m_d)\,\bar s d \\
 &\quad+\sum_j C_{ij}^{(6),q}(\Lambda,m_q)\,\mathcal Q_j^{(6)}+\cdots .
 \end{aligned}
 \label{eq:pure-loop-ope}
\end{equation}
Here, $\Lambda$ denotes the UV cutoff, set by $\Lambda\sim a^{-1}$ on the lattice. The dimension-six operators $\mathcal Q_j^{(6)}$ have the same dimension as the current-current operators in the weak Lagrangian, while the ellipsis denotes higher-dimensional operators.

The flavor-changing vector Ward identity gives
\begin{equation}
(m_s-m_d)\bar{s}d=\partial_\mu(\bar{s}\gamma_\mu d),
\end{equation}
up to convention-dependent signs. Thus, for $m_s\ne m_d$, the mass-dressed scalar operator can be expressed as
\begin{equation}
(m_s+m_d)\bar{s}d
=\frac{m_s+m_d}{m_s-m_d}
\partial_\mu(\bar{s}\gamma_\mu d).
\end{equation} As shown in Ref.~\cite{Isidori:2005tv}, this contribution does not affect the on-shell physical $K\to\pi\ell^+\ell^-$ amplitude. The leading lower-dimensional mixing therefore drops out of the physical amplitude, leaving dimension-six operators as the leading surviving short-distance contribution. Their ordinary same-dimension renormalization is at most logarithmically divergent. After the $u-c$ GIM subtraction, the UV divergence is regulated by the charm-quark mass scale, rendering the short-distance contribution finite while leaving a logarithmic dependence on the charm-quark mass.

\subsection{UV behavior of the loop-in contraction}

We analyze the UV behavior of the loop-in contraction using an operator-product expansion (OPE) for the terms relevant to the $K\to\pi$ matrix element. 
We use the local weak Lagrangian density $\Lweak_E$ with the convention defined in the main text, and
\begin{equation}
 \begin{aligned}
 J_\nu^E(x_E)&=\bar q(x_E)\gamma_\nu^Eq(x_E),
 \qquad x_E=(t_E,\vec{x}).
 \end{aligned}
 \label{eq:SAlt1_definitions}
\end{equation}
The Euclidean OPE takes the form
\begin{equation}
 T\!\left\{J_\nu^E(x_E)Q_{i,\rm GIM}(0)\right\}
 =\sum_{d,j}C_{\nu;i,j}^{(d)}(x_E;\mu)\,
 \mathcal O_j^{(d)}(0;\mu),
 \label{eq:SAlt2_ope}
\end{equation}
where $Q_{i,\rm GIM}=Q_i^u-Q_i^c$, $\dim J_\nu^E=3$, $\dim Q_{i,\rm GIM}=6$, so that
\begin{equation}
 \dim C_{\nu;i,j}^{(d)}=9-d.
 \label{eq:SAlt3_coefficient_dimension}
\end{equation}
For operators with $d\ge4$ and $r=|x_E|\to0$, the coefficient is therefore no more singular than
\begin{equation}
 C_{\nu;i,j}^{(d)}(x_E)=O\!\left(r^{d-9}\right)
 \le O(r^{-5}).
 \label{eq:SAlt5_higher_dimensional_coefficient}
\end{equation}

On the lattice, the short-distance singularities of the OPE coefficients are regulated by the finite lattice spacing $a$. For the conserved lattice vector current, the subtraction of the constant term in the Fourier kernel is justified by the exact lattice Ward identity and leaves the physical amplitude unchanged. For the local vector current, however, the corresponding Ward identity does not hold exactly at finite lattice spacing, even after renormalization. Consequently, the unsubtracted amplitude may contain additional short-distance contact terms, including finite contributions that survive the GIM subtraction.

The crucial observation is that momentum-independent contact terms are eliminated by the subtracted Fourier kernel, irrespective of whether the current is conserved. We show below, using the OPE, that the remaining momentum-dependent contributions are UV finite after the GIM subtraction. Since the renormalized local and conserved currents have the same continuum limit, and the subtracted amplitude is free of UV divergences, the two formulations yield the same physical result in the continuum limit.

Multiplying by the subtracted Fourier kernel
\begin{equation}
K(q,x_E)=e^{q^0t_E}e^{-i\vec q\cdot\vec x}-1=O(r),
\end{equation}
and integrating over a small four-dimensional ball, the most singular $d=4$ contribution scales as
\begin{equation}
 \int_{r<\rho}d^4x_E\,K(q,x_E)C_{\nu;i,j}^{(4)}(x_E)
 \sim r^3dr\,r\,r^{-5}=\frac{dr}{r}.
 \label{eq:SAlt6_dimension_four_counting}
\end{equation}
The leading, mass-independent coefficient is flavor blind and hence cancels between the $u$- and $c$-quark contributions in the GIM combination. 
The remaining terms are less singular. Thus, all contributions from operators with $d\ge4$ are UV finite after the GIM subtraction.

For $d=3$, the bilinears that can have nonvanishing matrix elements between the pseudoscalar states are
\begin{equation}
 \mathcal O_V^\nu=\bar s\gamma^\nu d,
 \qquad
 \mathcal O_S=\bar s d,
 \qquad
 \mathcal O_T^{\nu\rho}=\bar s\sigma^{\nu\rho}d.
 \label{eq:SAlt7_surviving_operators}
\end{equation}
The axial-vector and pseudoscalar bilinears have vanishing $K\to\pi$ matrix elements by parity. 
At short distances, in the chiral limit, the scalar and tensor bilinears are chirality flipping, whereas the leading $V-A$ structure is chirality preserving. Chiral symmetry therefore requires an additional factor of $m_s$ or $m_d$ for the scalar and tensor contributions. They consequently behave as effective dimension-4 operators and are UV finite by the power counting above, together with the GIM subtraction.

The vector bilinear is thus the only dimension-three contribution requiring separate consideration. We first consider the Wilson coefficient at nonzero separation, leaving possible contact terms localized at $x_E=0$ to be treated separately. Euclidean $O(4)$ symmetry constrains the coefficient to the form
\begin{equation}
C_{\nu\mu}^{q,V}(x_E)
=A_q(x_E^2)\delta_{\nu\mu}
+B_q(x_E^2)x_{E\nu}x_{E\mu},
\label{eq}
\end{equation}
where the index $\nu$ originates from the electromagnetic current $J_\nu^E$, while $\mu$ is associated with the vector bilinear $\mathcal O_V^\mu=\bar{s}\gamma^\mu d$. Dimensional analysis gives $A_q=O(r^{-6})$ and $B_q=O(r^{-8})$, up to logarithmic corrections.
For the symmetric ball $B_\rho=\{x_E:r<\rho\}$, define
\begin{equation}
 I^{q,V}_{\nu\mu}(\rho)
 =\int_{B_\rho}d^4x_E\,K(q,x_E)C^{q,V}_{\nu\mu}(x_E).
 \label{eq:SAlt11_vector_integral}
\end{equation}
Since $C^{q,V}_{\nu\mu}(x_E)$ is even under $x_E\to-x_E$, the integral can be written as
\begin{equation}
 I^{q,V}_{\nu\mu}(\rho)
 =\frac12\int_{B_\rho}d^4x_E\,C^{q,V}_{\nu\mu}(x_E)
 \left[K(q,x_E)+K(q,-x_E)\right].
 \label{eq:SAlt12_vector_pairing}
\end{equation}
The corresponding even kernel is
\begin{equation}
\begin{aligned}
 K_{\rm even}(q,x_E)
 &\equiv\frac12\left[K(q,x_E)+K(q,-x_E)\right] \\
 &=\cosh(q^0t_E)\cos(\vec q\cdot\vec x_E)-1\\
 &\quad-i\sinh(q^0t_E)\sin(\vec q\cdot\vec x_E)\\
 &=O(r^2).
\end{aligned}
 \label{eq:SAlt13_even_kernel}
\end{equation}
The resulting short-distance behavior is
\begin{equation}
 \int_{r<\rho}d^4x_E\,K_{\rm even}(q,x_E)C^{q,V}_{\nu\mu}(x_E)
 \sim r^3dr\,r^2r^{-6}=\frac{dr}{r}.
 \label{eq:SAlt15_vector_counting}
\end{equation}
The Fourier subtraction therefore reduces the quadratic divergence of the dimension-3 vector contribution to a logarithmic divergence. 
The mass-independent part of this logarithmic divergence cancels in the GIM combination,
rendering the loop-in contribution UV finite. 
Contact terms proportional to $\delta^{(4)}(x_E)$, which are independent of the Fourier momentum, are eliminated by the subtracted kernel since $K(q,0)=0$. Possible derivative contact terms are constrained by the same short-distance power counting and GIM subtraction discussed above. 
Higher-order QCD corrections may generate contributions depending on both the photon momentum and the internal quark mass. However, the mass-dependent terms in the short-distance OPE coefficients are less singular and, when combined with the subtracted Fourier kernel, give UV-finite contributions.
Thus, the local-current formulation yields the same continuum limit as the conserved-current formulation.

\subsection{Direct charm-quark mass extrapolation of $V(z)$}
\label{sec:direct-charm-vz}

As discussed in the analysis of the UV behavior of the pure-loop and loop-in contractions, after the GIM subtraction the residual short-distance dependence on the charm-quark mass is logarithmic. 
Motivated by this behavior, we parameterize the charm-mass dependence of $V(z)$ at each momentum and on each ensemble as
\begin{equation}
 V(z,x)=A(z)+B(z)\log x,\qquad
 x=\frac{M_{\eta_c}}{M_{\eta_c}^{\rm phys}},
 \label{eq:direct-charm-vz}
\end{equation}
where $M_{\eta_c}^{\rm phys}=2.9839\,\GeV$ and $A(z)=V(z,1)$ gives the estimate at the physical charm-quark mass. We use $M_{\eta_c}$ as a proxy for the charm-quark mass. 
The simulated values span $M_{\eta_c}=0.7$--$1.1\,\GeV$ on 24D and $0.7$--$1.3\,\GeV$ on 32Dfine. We retain only the leading logarithmic dependence, motivated by the expected short-distance behavior and the smooth charm-mass dependence observed in the data. The results at different charm-quark masses are strongly correlated, and their individual statistical uncertainties therefore do not directly reflect the uncertainty in the charm-mass dependence.

As a nontrivial test of this ansatz, we perform an out-of-sample prediction on the 32Dfine ensemble. At each momentum, we fit the results at $\log x=-1.45$, $-1.20$, and $-1.00$, corresponding to $M_{\eta_c}=0.7$, $0.9$, and $1.1\,\GeV$, respectively, and predict the value at $\log x=-0.83$ ($M_{\eta_c}=1.3\,\GeV$), which is excluded from the fit. Table~\ref{tab:32dfine-charm-holdout} compares these predictions with the direct lattice results. The agreement across all six kinematic points supports the logarithmic ansatz over the simulated charm-mass range.

\begin{table}[htb]
\centering
\begin{tabular}{cccccc}
\hline\hline
Kinematics & Mode & $z$ & $V_{\rm pred}$ & $V_{\rm meas}$ & $\delta$ \\
\hline
$K(\vecp 0)\to\pi(\vecp p)$ & 1 & $-0.163$ & $-0.87(26)$ & $-0.85(26)$ & $0.013(22)$ \\
 & 2 & $-0.581$ & $-0.76(32)$ & $-0.76(32)$ & $-0.006(23)$ \\
 & 3 & $-0.910$ & $-1.26(39)$ & $-1.27(39)$ & $-0.007(23)$ \\
\hline
$K(\vecp k)\to\pi(\vecp 0)$ & 1 & $0.414$ & $-0.51(50)$ & $-0.45(50)$ & $0.14(19)$ \\
 & 2 & $0.340$ & $-0.69(42)$ & $-0.71(42)$ & $-0.020(32)$ \\
 & 3 & $0.279$ & $0.05(63)$ & $0.08(63)$ & $-0.4(50.0)$ \\
\hline\hline
\end{tabular}
\caption{Comparison between the extrapolated and directly calculated values of $V(z)$ at $\log x=-0.83$ on the 32Dfine ensemble. Mode 1, 2, 3 correspond to $|\vecp p|$ or $|\vec{k}|=(1,\sqrt{2},\sqrt{3})\,2\pi/L$. The extrapolation uses the three points at $\log x=-1.45$, $-1.20$, and $-1.00$. The relative difference is $\delta=(V_{\rm pred}-V_{\rm meas})/V_{\rm meas}$.}
\label{tab:32dfine-charm-holdout}
\end{table}

\begin{figure}[htb]
 \centering
 \includegraphics[width=\columnwidth]{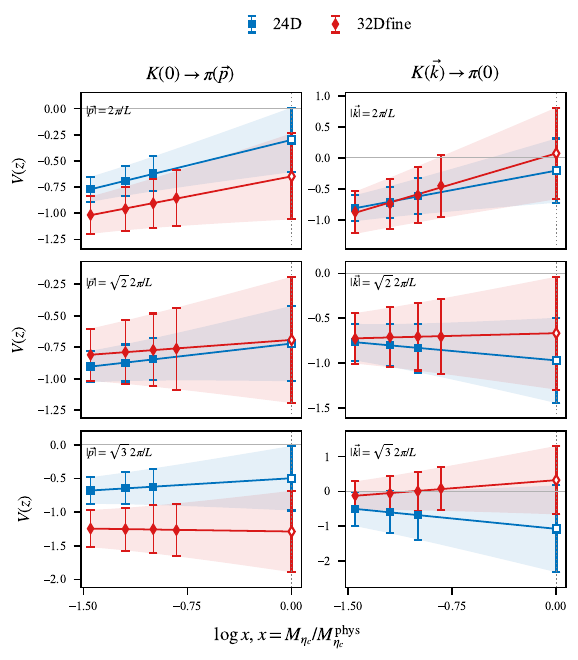}
 \caption{Charm-mass dependence of $V(z)$ for $K(\vecp{0})\to\pi(\vecp{p})$ and $K(\vecp{k})\to\pi(\vecp{0})$, obtained using Method 3 and Method 2, respectively. The bands show fits of the form $A+B\log(M_{\eta_c}/M_{\eta_c}^{\rm phys})$ with statistical uncertainties, and the rightmost points indicate the extrapolated values at the physical charm-quark mass.}
 \label{fig:direct-charm-vz}
\end{figure}

\begin{table}[htb]
\setlength{\tabcolsep}{1.8pt}
\begin{ruledtabular}
\fontsize{8}{9.5}\selectfont
\begin{tabular}{c c c c}
$n$ & Ensemble & Intercept $A(z)$ & Slope $B(z)$ \\
\hline
\multicolumn{4}{c}{$K(\vecp{0})\to\pi(\vecp{p})$} \\
1 & 24D & $-0.30(31)$ & $0.33(16)$ \\
 & 32Dfine & $-0.65(41)$ & $0.26(20)$ \\
2 & 24D & $-0.72(30)$ & $0.13(15)$ \\
 & 32Dfine & $-0.69(50)$ & $0.08(23)$ \\
3 & 24D & $-0.50(48)$ & $0.13(23)$ \\
 & 32Dfine & $-1.29(60)$ & $-0.03(30)$ \\
\hline
\multicolumn{4}{c}{$K(\vecp{k})\to\pi(\vecp{0})$} \\
1 & 24D & $-0.21(52)$ & $0.42(26)$ \\
 & 32Dfine & $0.07(74)$ & $0.66(32)$ \\
2 & 24D & $-0.97(48)$ & $-0.14(24)$ \\
 & 32Dfine & $-0.67(63)$ & $0.04(28)$ \\
3 & 24D & $-1.1(1.2)$ & $-0.40(61)$ \\
 & 32Dfine & $0.32(99)$ & $0.31(48)$ \\
\end{tabular}
\end{ruledtabular}
\caption{Results of the logarithmic fits $V(z,x)=A(z)+B(z)\log x$ with $x=M_{\eta_c}/M_{\eta_c}^{\rm phys}$ shown in Fig.~\ref{fig:direct-charm-vz}. The momentum mode is labeled by $n=|\vecp p|^2/(2\pi/L)^2$ for the spacelike kinematics and $n=|\vecp k|^2/(2\pi/L)^2$ for the timelike kinematics. All uncertainties are statistical.}
\label{tab:direct-charm-log-fits}
\end{table}

For our final determination, we perform the logarithmic charm-mass extrapolation using all available data points: three on the 24D ensemble and four on the 32Dfine ensemble. Fig.~\ref{fig:direct-charm-vz} shows the charm-mass dependence and the extrapolation to the physical charm-quark mass, with the corresponding fit parameters summarized in Table~\ref{tab:direct-charm-log-fits}. The extrapolation to $\log x=0$ substantially increases the statistical uncertainty, while the shift in the central value from the simulated charm-mass region to the physical point remains smaller than the extrapolated uncertainty. The physical-charm results are shown in the lower panel of Fig.~\ref{fig:paired-v-experiment}. Despite the increased uncertainties, the overall pattern of predominantly negative $V(z)$ values remains unchanged after the extrapolation.

\subsection{Fits to the $z$ dependence of $V(z)$}

To assess the sensitivity to the assumed $z$ dependence, we fit $V(z)$ at $M_{\eta_c}=1.1\,\GeV$ using polynomial parameterizations up to quadratic order, both including and excluding the phenomenological contribution $V^{\pi\pi}(z)$. The resulting fit parameters, values of $V(0)$, and $\chi^2/\mathrm{dof}$ are summarized in Table~\ref{tab:fit}. The corresponding results after extrapolation to the physical charm-quark mass, using $M_{\eta_c}=M_{\eta_c}^{\rm phys}$, are given in Table~\ref{tab:physical-charm-vz-fits}. 

\begin{table}[htb]
\setlength{\tabcolsep}{1.8pt}
\begin{ruledtabular}
\fontsize{8}{9.5}\selectfont
\begin{tabular}{c c c c c c c}
Fit & Ensemble & $a_+$ & $b_+$ & $c_+$ & $V(0)$
& $\chi^2/\mathrm{dof}$ \\
\hline
\multicolumn{7}{c}{$V^{\pi\pi}(z)\ \mathrm{included}$} \\
$O(z^0)$ & 24D
& $-0.67(11)$ & -- & -- & $-0.67(11)$ & $0.37$ \\
 & 32Df
& $-0.77(19)$ & -- & -- & $-0.77(19)$ & $0.99$ \\

$O(z^1)$ & 24D
& $-0.64(11)$ & $0.17(26)$ & -- & $-0.64(11)$ & $0.34$ \\
 & 32Df
& $-0.66(20)$ & $0.52(41)$ & -- & $-0.66(20)$ & $0.55$ \\

$O(z^2)$ & 24D
& $-0.69(15)$ & $0.27(41)$ & $0.33(72)$ & $-0.69(15)$ & $0.36$ \\
 & 32Df
& $-0.68(19)$ & $0.59(75)$ & $0.17(88)$ & $-0.68(19)$ & $0.72$ \\

\hline
\multicolumn{7}{c}{$V^{\pi\pi}(z)\ \mathrm{excluded}$} \\
$O(z^0)$ & 24D
& $-0.72(11)$ & -- & -- & $-0.72(11)$ & $0.27$ \\
 & 32Df
& $-0.82(19)$ & -- & -- & $-0.82(19)$ & $0.82$ \\

$O(z^1)$ & 24D
& $-0.71(11)$ & $0.05(26)$ & -- & $-0.71(11)$ &$0.32$ \\
 & 32Df
& $-0.72(20)$ & $0.44(41)$ & -- & $-0.72(20)$ & $0.54$ \\

$O(z^2)$ & 24D
& $-0.71(15)$ & $0.05(41)$ & $0.02(72)$ & $-0.71(15)$ & $0.43$ \\
 & 32Df
& $-0.71(19)$ & $0.40(75)$ & $-0.08(88)$ & $-0.71(19)$ & $0.71$ \\

\end{tabular}
\end{ruledtabular}
\caption{%
Results of fits to the lattice data for $V(z)$ using polynomial
parameterizations of increasing order. 
The charm-quark mass on both ensembles corresponds to $M_{\eta_c}=1.1\,\GeV$.
For the first three rows,
the fit ansatz is
$V(z)=a_+ + b_+z + c_+z^2 + V^{\pi\pi}(z)$,
truncated at order $O(z^0)$, $O(z^1)$, and $O(z^2)$, respectively.
For the last three rows, the phenomenological input
$V^{\pi\pi}(z)$ is set to zero. The corresponding fit ansatzes are
$V(z)=a_+$, $a_+ + b_+z$, and
$a_+ + b_+z+c_+z^2$, respectively.
}
\label{tab:fit}
\end{table}

\begin{table}[htb]
\setlength{\tabcolsep}{1.8pt}
\begin{ruledtabular}
\fontsize{8}{9.5}\selectfont
\begin{tabular}{c c c c c c c}
Fit & Ensemble & $a_+$ & $b_+$ & $c_+$ & $V(0)$ & $\chi^2/\mathrm{dof}$ \\
\hline
\multicolumn{7}{c}{$V^{\pi\pi}(z)\ \mathrm{included}$} \\
$O(z^0)$ & 24D & $-0.50(19)$ & -- & -- & $-0.50(19)$ & $0.49$ \\
 & 32Df & $-0.58(35)$ & -- & -- & $-0.58(35)$ & $0.70$ \\
$O(z^1)$ & 24D & $-0.47(20)$ & $0.17(45)$ & -- & $-0.47(20)$ & $0.57$ \\
 & 32Df & $-0.41(38)$ & $0.79(68)$ & -- & $-0.41(38)$ & $0.31$ \\
$O(z^2)$ & 24D & $-0.47(28)$ & $0.17(74)$ & $0.0(1.4)$ & $-0.47(28)$ & $0.75$ \\
 & 32Df & $-0.42(37)$ & $0.8(1.2)$ & $0.1(1.4)$ & $-0.42(37)$ & $0.41$ \\
\hline
\multicolumn{7}{c}{$V^{\pi\pi}(z)\ \mathrm{excluded}$} \\
$O(z^0)$ & 24D & $-0.55(19)$ & -- & -- & $-0.55(19)$ & $0.48$ \\
 & 32Df & $-0.63(35)$ & -- & -- & $-0.63(35)$ & $0.61$ \\
$O(z^1)$ & 24D & $-0.54(20)$ & $0.05(45)$ & -- & $-0.54(20)$ & $0.60$ \\
 & 32Df & $-0.48(38)$ & $0.71(68)$ & -- & $-0.48(38)$ & $0.30$ \\
$O(z^2)$ & 24D & $-0.50(28)$ & $-0.04(74)$ & $-0.3(1.4)$ & $-0.50(28)$ & $0.77$ \\
 & 32Df & $-0.45(37)$ & $0.6(1.2)$ & $-0.2(1.4)$ & $-0.45(37)$ & $0.40$ \\
\end{tabular}
\end{ruledtabular}
\caption{Fit results corresponding to Table~\ref{tab:fit} after extrapolation to $M_{\eta_c}=M_{\eta_c}^{\rm phys}$.}
\label{tab:physical-charm-vz-fits}
\end{table}

The sign of $V(0)$ is stable against the fit order, the inclusion or exclusion of $V^{\pi\pi}(z)$, and the extrapolation to the physical charm-quark mass.

\subsection{Three-pion intermediate state}

In this section, we use leading-order chiral perturbation theory to estimate the contributions from three-pion intermediate states. Our aim is to show that the exponentially growing terms associated with three-pion states below the kaon energy, together with the corresponding power-law finite-volume effects, are negligible at the present level of precision.

Considering the Euclidean amplitude $M(t_E,\vec k,\vec p)$ with $\vec{k}=\vec{0}$ and inserting finite-volume energy eigenstates between the two insertions, the
negative-time ($t_E<0$) contribution reads
\begin{equation}
 M(t_E,\vec 0,\vec p)\supset\sum_n A_n(\vec p)
 \exp\!\left[-(M_K-E_n)t_E\right],
 \label{eq:S_three_pion_spectral}
\end{equation}
where $M_K$ is the kaon mass, $E_n$ the energy of the $n$th finite-volume
state, and $A_n(\vecp p)$ its overlap.  The one-pion state is removed
in the main analysis.  The two-pion intermediate state does not contribute to the rare-decay amplitude~\cite{Christ:2015aha}.  The lowest remaining
subthreshold multiparticle state is therefore the three-pion threshold state.
Neglecting its the energy shift due to interactions,
\begin{equation}
 E_{3\pi}^{\rm th}=3M_\pi,
 \qquad \Delta_{3\pi}=M_K-3M_\pi ,
 \label{eq:S_three_pion_gap}
\end{equation}
where $M_\pi$ is the pion mass and $\Delta_{3\pi}$ the energy gap to the
kaon.

\begin{figure}[htb]
 \centering
 \includegraphics[width=\columnwidth]{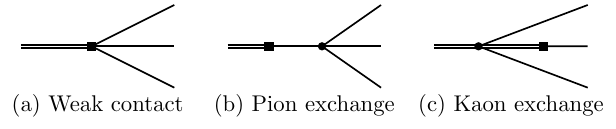}
 \caption{Leading-order diagrams contributing to $K^+\to3\pi$: (a) weak contact, (b) pion exchange and (c) kaon exchange. Squares denote weak vertices and circles strong vertices. The double line identifies the kaon and single line for pion here.}
 \label{fig:three-pion-weak-diagrams}
\end{figure}

\paragraph{\textit{(a) \(K\to3\pi\) matrix element.}}
Let $\mathcal W_c=\langle3\pi,c|\Lweak_E|K^+\rangle$ be the weak matrix
element, and define
\begin{equation}
 g\equiv g_8+\frac23g_{27}.
\end{equation}
Here $g_8$ and $g_{27}$ are the dimensionless octet and 27-plet weak couplings, while $g'_8$ multiplies the $O(p^2)$ weak-mass operator in the $\Delta S=1$ chiral Lagrangian. This operator involves the quark-mass spurion and has the same chiral transformation structure as the scalar density $\bar s d$. All weak matrix elements
below carry the normalization
$\mathcal N_W=G_F|V_{ud}V_{us}|/\sqrt2$, where $G_F$ is the Fermi constant
and $V_{ud},V_{us}$ are Cabibbo--Kobayashi--Maskawa matrix elements.  The LO
calculation contains the weak contact tree and the pion- and kaon-exchange
trees of Fig.~\ref{fig:three-pion-weak-diagrams}. We evaluate these amplitudes with the kaon and the three-pion intermediate state both at zero total spatial momentum. The weak-mass operator contributes to the individual tree diagrams, so we retain the terms proportional to $g'_8$:
\begin{align}
 \mathcal W_{+00}^{(a)}
 &=\mathcal N_Wg\,\frac{2M_\pi(M_K+M_\pi)}{3}
 -\frac13\mathcal N_Wg'_8M_K^2,\\
 \mathcal W_{+00}^{(b)}
 &=\mathcal N_Wg\,\frac{M_\pi M_K}{4}
 -\frac1{12}\mathcal N_Wg'_8M_K^2,\\
 \mathcal W_{+00}^{(c)}
 &=-\mathcal N_Wg\,
 \frac{M_\pi(M_K^2-4M_\pi^2)}{6(M_K-M_\pi)}\nonumber\\
 &\quad+\frac{\mathcal N_Wg'_8M_K^2(M_K+2M_\pi)}
       {6(M_K-M_\pi)}.
 \label{eq:S_three_pion_weak_diagrams}
\end{align}
The three contributions correspond to the (a), (b), (c) diagrams in Fig.~\ref{fig:three-pion-weak-diagrams} respectively. Their sum is
\begin{align}
 \mathcal W_{+00}^{\rm LO}(g'_8)
 &=\mathcal N_Wg\,
 \frac{M_\pi M_K(3M_K-M_\pi)}{4(M_K-M_\pi)}\nonumber\\
 &\quad-\mathcal N_Wg'_8M_K^2
 \frac{M_K-3M_\pi}{4(M_K-M_\pi)},
 \label{eq:S_three_pion_weak_total}\\
 \mathcal W_{++-}^{\rm LO}(g'_8)
 &=2\mathcal W_{+00}^{\rm LO}(g'_8).
\end{align}
Subtraction Method 3 described above replaces $\Lweak_E$ by $\Lweak_E-c\,\bar s d$, where
$\bar s d$ is the scalar density and
$c=\langle\pi|\Lweak_E|K\rangle/\langle\pi|\bar s d|K\rangle$, so as to remove
the one-pion state.  The scalar-density ratio and the resulting subtracted
weak amplitudes are
\begin{align}
 \rho_{+00}
 &=\frac{\langle3\pi,+00|\bar s d|K^+\rangle}
           {\langle\pi^+|\bar s d|K^+\rangle}
 =-\frac{M_K-3M_\pi}{8F_0^2(M_K-M_\pi)},\\
 \mathcal W_{+00}^{\prime\,{\rm LO}}
 &=\mathcal N_Wg\,
 \frac{M_\pi M_K(M_K+M_\pi)}{2(M_K-M_\pi)},\\
 \mathcal W_{++-}^{\prime\,{\rm LO}}
 &=2\mathcal W_{+00}^{\prime\,{\rm LO}}.
 \label{eq:S_three_pion_method2_shift}
\end{align}
The scalar-density subtraction in Method 3 also removes the weak-mass contribution to the three-pion amplitude at LO. Specifically, the combination $\mathcal W_{+00}^{\rm LO}(g'_8)-\rho_{+00}\langle\pi|\Lweak_E|K\rangle$ is independent of $g'_8$, as reflected in the subtracted amplitudes $\mathcal W_{+00}^{\prime,{\rm LO}}$ and $\mathcal W_{++-}^{\prime,{\rm LO}}$ in Eq.~\eqref{eq:S_three_pion_method2_shift}.

\begin{figure}[t]
 \centering
 \includegraphics[width=\columnwidth]{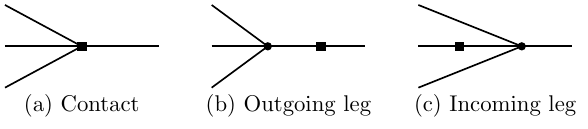}
 \caption{Leading-order diagrams contributing to $\langle\pi^+(p)|J_0^E|3\pi,c\rangle$: (a) the contact term and (b,c) the charged-pion pole terms. Squares denote current insertions and circles denote strong-interaction vertices.}
 \label{fig:three-pion-current-diagrams}
\end{figure}

\paragraph{\textit{(b) \(\pi\text{-}J\text{-}3\pi\) matrix element.}}
For both channels $c=\pi^+\pi^0\pi^0$ or $c=\pi^+\pi^+\pi^-$, we label the momentum of the non-identical pion by
$p_1$ and those of the identical pair by $p_2,p_3$, and define
$H_{c,\mu}\equiv\langle\pi^+(p)|J_\mu^E|3\pi,c\rangle$.  At threshold the momenta of the three pions are each $p_i=(M_\pi,\vecp{0})$, while
the external pion carries $p=(E_\pi,\vecp p)$ with
$E_\pi=\sqrt{M_\pi^2+|\vecp p|^2}$.  The leading-order (LO) calculation
contains one direct contact tree built from the four-pion component
$J^{(4)}$ of the current, and two pole trees in which the quadratic (two-pion)
component $J^{(2)}$ attaches to a charged pion, as shown in
Fig.~\ref{fig:three-pion-current-diagrams}.  Throughout, $F_0=f_\pi/\sqrt{2}$, where the decay constant $f_\pi\simeq 130$ MeV. The temporal components of the three diagrams are
\begin{align}
 H_{+00,0}^{(a)}
 &=-\frac{2(E_\pi+M_\pi)}{3F_0^2},\\
 H_{+00,0}^{(b)}
 &=-\frac{E_\pi+3M_\pi}{24F_0^2},\\
 H_{+00,0}^{(c)}
 &=\frac{4E_\pi+5M_\pi}{12F_0^2},
 \label{eq:S_three_pion_current_diagrams}
\end{align}
where the superscripts $(a)$, $(b)$ and $(c)$ correspond to the three diagrams in Fig.~\ref{fig:three-pion-current-diagrams} respectively. Their sum gives the matrix
element entering the spectral sum:
\begin{equation}
 H_{+00,0}^{\rm LO}
 =-\frac{3(E_\pi+M_\pi)}{8F_0^2},
 \qquad
 H_{++-,0}^{\rm LO}=2H_{+00,0}^{\rm LO},
 \label{eq:S_three_pion_current_total}
\end{equation}
the second relation following from isospin symmetry.

\paragraph{\textit{(c) Effect on \(V(z)\).}}
Applying the Method 3 subtraction to the coefficient in 
Eq.~\eqref{eq:S_three_pion_method2_shift} and inserting the resulting subtracted coefficient into the completeness relation saturated by the three-pion threshold state, with the $1/2!$ Bose factor for the identical pair included, we obtain the finite-volume coefficient
\begin{equation}
 \begin{split}
  A_{3\pi}^{\rm LO,FV}
  &=-\frac{15\mathcal N_WM_K(M_K+M_\pi)}
        {256M_\pi^2F_0^2L^6(M_K-M_\pi)}\\
  &\qquad\times(E_\pi+M_\pi)
  \left(g_8+\frac23g_{27}\right),
 \end{split}
 \label{eq:S_three_pion_A_LO}
\end{equation}
where $L$ is the spatial extent of the lattice.  This has mass dimension two,
as required for $M(t_E,\vec k,\vecp p)$.

\begin{figure}[b]
 \centering
 \includegraphics[width=\columnwidth]{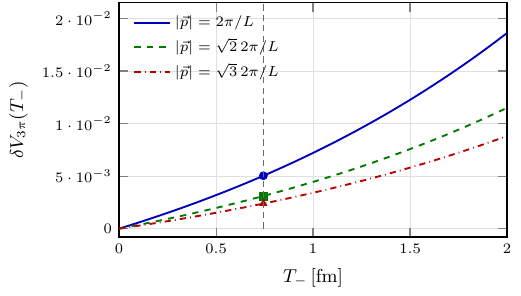}
 \caption{Cumulative leading-order three-pion threshold contribution for momentum modes $|\vecp{p}|=(1,\sqrt{2},\sqrt{3} )2\pi/L$ on the 24D ensemble. The dashed line and filled symbols indicate $T_-=0.744\,\fm$.}
 \label{fig:three-pion-vz-cut}
\end{figure}

For a numerical benchmark, we take $g_8=3.61$ and $g_{27}=0.297$, using values determined from 
a next-to-leading-order analysis of $K\to2\pi$ decays including isospin-violating effects~\cite{Cirigliano:2011ny}. 
We further use $f_\pi=0.1302\,\GeV$, $\mathcal N_W=1.8013\times10^{-6}\,\GeV^{-2}$, and the isospin-symmetric meson masses $M_\pi=0.1350\,\GeV$ and $M_K=0.4957\,\GeV$~\cite{RBC:2014ntl}.
We use
$E_\pi=\sqrt{M_\pi^2+\vecpp p^{2}}$, where
$\vec p=\frac{2\pi}{L}\vec n$ and $\vec n$ is an integer-valued momentum vector. The integrated negative-time contribution is
\begin{equation}
 \begin{aligned}
  &M_{3\pi}(t_E,\vec0,\vecp p)=A_{3\pi}(\vecp p)
  e^{\Delta_{3\pi}|t_E|},\qquad t_E<0,\\
  &\delta\mathcal A^{\rm E}_{3\pi}(T_-)
  \equiv\int_{-T_-}^{0}\hspace{-10pt}dt_E\,M_{3\pi}(t_E,\vec0,\vecp p)
  =A_{3\pi}(\vecp p)\,
  \frac{e^{\Delta_{3\pi}T_-}-1}{\Delta_{3\pi}}.
 \end{aligned}
\label{eq:S_three_pion_integral}
\end{equation}

To express this result in terms of the contribution of the three-pion intermediate state to $V(z)$ as defined in Eq.~\eqref{eq:minkowski_Vz}, we define
\begin{eqnarray}
  C_0(\vecp p)&\equiv& q^2(M_K+E_\pi)
  -(M_K^2-M_\pi^2)(M_K-E_\pi)\nonumber\\
  &=&-2M_K|\vecp p|^2<0,
 \label{eq:S_three_pion_C0}
\end{eqnarray}
where $k=(M_K,\vecp{0})$ and $p=(E_\pi,\vecp{p})$ are the kaon and pion
four-momenta and $q=k-p$.  Matching $\mathcal A_0=-i\mathcal A_0^{E}$ to
Eq.~\eqref{eq:minkowski_Vz} then gives the cumulative three-pion
contribution to the form factor $V(z)$ defined in the main text,
\begin{equation}
 \delta V_{3\pi}(T_-;\vecp p)=
 \frac{(4\pi)^2}{G_F C_0(\vecp p)}
 A_{3\pi}^{\rm LO,FV}(\vecp p)\,
 \frac{e^{\Delta_{3\pi}T_-}-1}{\Delta_{3\pi}}.
 \label{eq:S_three_pion_Vz_cut}
\end{equation}

The cumulative three-pion contribution as a function of $T_-$ is shown in Fig.~\ref{fig:three-pion-vz-cut} for three momentum modes on the 24D ensemble. At the adopted integration limit $T_-=0.744,\fm$, the estimated contribution is $|\delta V_{3\pi}|\lesssim5.3\times10^{-3}$, consistent with the expected phase-space suppression of the associated exponentially growing term. Although the discussion above assumes the kinematics $\vec k=\vec0$, $\vec p\neq\vec0$, the same conclusion applies to $\vec k\neq\vec0$, $\vec p=\vec0$. This also implies that the associated three-pion finite-volume effects are negligible, since only three-pion intermediate states with energies below $M_K$ can generate the relevant power-law finite-volume effects.

\end{document}